# Multi-scale Metabolic Modeling and Simulation

**Peter E. Carstensen[a], Teddy Groves[b], Lars K. Nielsen[b], Ulrich Krühne[c], Krist V. Gernaey[c], and John B. Jørgensen[a]***

[a] Technical University of Denmark, Department of Applied Mathematics and Computer Science, Kgs. Lyngby, Denmark
[b] Technical University of Denmark, Novo Nordisk Foundation Biotechnology Research Institute for the Green Transition, Kgs. Lyngby, Denmark
[c] Technical University of Denmark, Department of Chemical and Biochemical Engineering, Kgs. Lyngby, Denmark

* Corresponding Author: jbjo@dtu.dk.

## ABSTRACT

Biological systems are governed by coupled interactions between intracellular metabolism and bioreactor operation that span multiple time scales. Constraint-based metabolic models are widely used to describe intracellular metabolism, but repeatedly solving the optimization problem at each time step in dynamic models introduces numerical challenges related to infeasibility and computational efficiency. This work presents a multi-scale modeling framework that integrates genome-scale, constraint-based metabolic models with dynamic bioreactor simulations. Intracellular metabolism is described using positive flux variables in a parsimonious flux balance analysis, and the resulting embedded optimization problem is replaced by a neural network surrogate. The surrogate provides a smooth approximation of the embedded optimization mapping and eliminates repeated linear program solves during simulation. The approach is demonstrated for fed-batch fermentation of *Escherichia coli*, in which the surrogate model yields intracellular fluxes under substrate-limited conditions, whereas the underlying linear program would otherwise be infeasible. The framework provides a continuous representation of intracellular metabolism suitable for dynamic simulation of genome-scale models in bioreactor configurations.

## INTRODUCTION

Biological systems are governed by coupled interactions between cellular metabolism and process operation. These interactions span multiple time scales, from the rapid kinetics of intracellular reactions to the slower dynamics of bioreactor processes. Enzymatic reactions typically occur on the order of milliseconds, transcription and translation evolve over seconds to minutes, and bioreactor-scale phenomena develop over hours to days [1]. Modeling such systems requires a consistent representation of processes across these time scales [1,2].

Constraint-based metabolic models are widely used to describe intracellular metabolism [3,4] and are commonly embedded in dynamic reactor simulations through dynamic flux balance analysis (dFBA) [5-7]. This formulation links genome-scale metabolism with extracellular mass balances. Numerical challenges arise when the optimization problem approaches infeasible regions, for example, when maintenance requirements cannot be satisfied [5, 6, 8]. Changes in active constraints induce a piecewise-defined dependence of intracellular fluxes on extracellular conditions, potentially leading to infeasibility in substrate-limited regions [9].

In this work, a multi-scale modeling framework is presented that integrates genome-scale, constraint-based metabolic models with dynamic bioreactor simulations by replacing the embedded optimization problem with a neural network surrogate [10]. The framework consists of three coupled components. A reactor model describing flow, mixing, and volume dynamics. An extracellular reaction model defining mass balances for substrates, biomass, and products, and a kinetic cell model providing intracellular metabolic fluxes as functions of the extracellular conditions, as shown in Figure 1. Before surrogate replacement, the multi-scale model is expressed as a coupled dynamic optimization problem. Extracellular states evolve according to dynamic mass balances, whereas intracellular fluxes are obtained by solving an optimization problem subject to stoichiometric constraints, flux bounds, and a biological objective. Intracellular metabolism is assumed to operate in a quasi-steady state relative to the reactor dynamics, reflecting the separation of time scales between fast metabolic reactions and slower process-level phenomena [1,5,7].

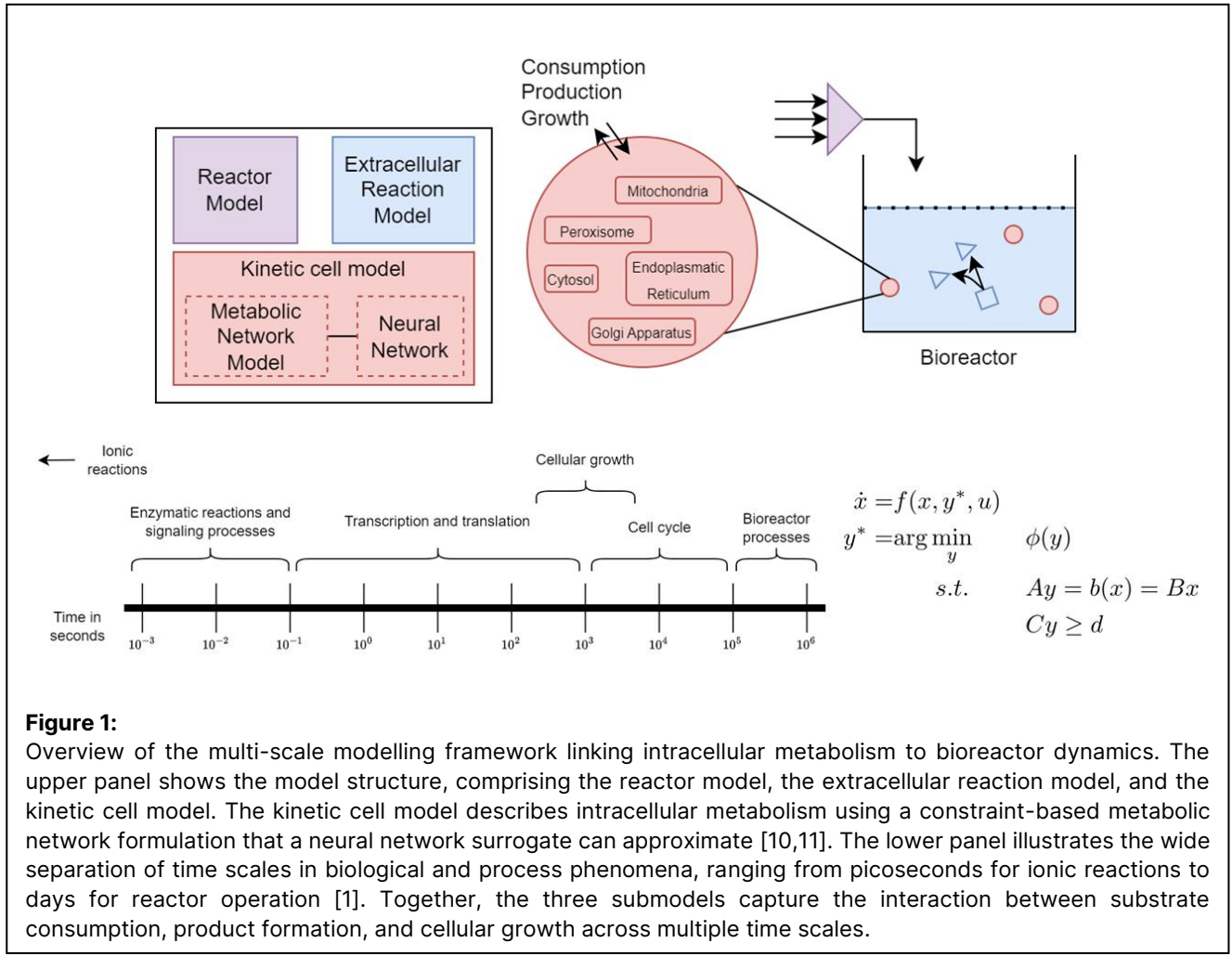

**Figure 1:**
Overview of the multi-scale modelling framework linking intracellular metabolism to bioreactor dynamics. The upper panel shows the model structure, comprising the reactor model, the extracellular reaction model, and the kinetic cell model. The kinetic cell model describes intracellular metabolism using a constraint-based metabolic network formulation that a neural network surrogate can approximate [10,11]. The lower panel illustrates the wide separation of time scales in biological and process phenomena, ranging from picoseconds for ionic reactions to days for reactor operation [1]. Together, the three submodels capture the interaction between substrate consumption, product formation, and cellular growth across multiple time scales.

The formulation is demonstrated using fed-batch fermentation of *Escherichia coli* with the genome-scale model iCH360 [3]. The coupled formulation links genome-scale metabolic models with reactor-scale dynamics within a single differentiable model.

The paper is structured as follows. The Method section presents the multi-scale modeling formulation, including the reactor model, extracellular mass balances, and kinetic cell model, as well as a description of the constraint-based metabolic formulation and the neural network surrogate model. The Results section presents simulation results for fed-batch fermentation of *E. coli* and a sensitivity analysis, followed by a discussion of model behavior and limitations, and a brief conclusion.

## METHOD

### Dynamic modeling

The multi-scale metabolic system is described by a set of ordinary differential equations coupled with optimal intracellular reaction rates. The extracellular dynamics depend on optimal rates obtained from an embedded optimization problem, or a surrogate model. The coupled system is written as,

$$\dot{x} = f(x, y^*, u) \qquad (1a)$$
$$y^*(x) = y^* = \arg\min_y \phi(y) \qquad (1b)$$
$$s.t. \quad Ay = b(x) = Bx \qquad (1c)$$
$$Cy \geq 0 \qquad (1d)$$

$x$ denotes the extracellular state variables, $y^*$ is the optimal rates, as a function of the extracellular conditions, and $u$ is the manipulated inputs. A surrogate representation may replace the embedded optimization problem in Equation (1b-1d) without altering the structure of the remaining submodels.

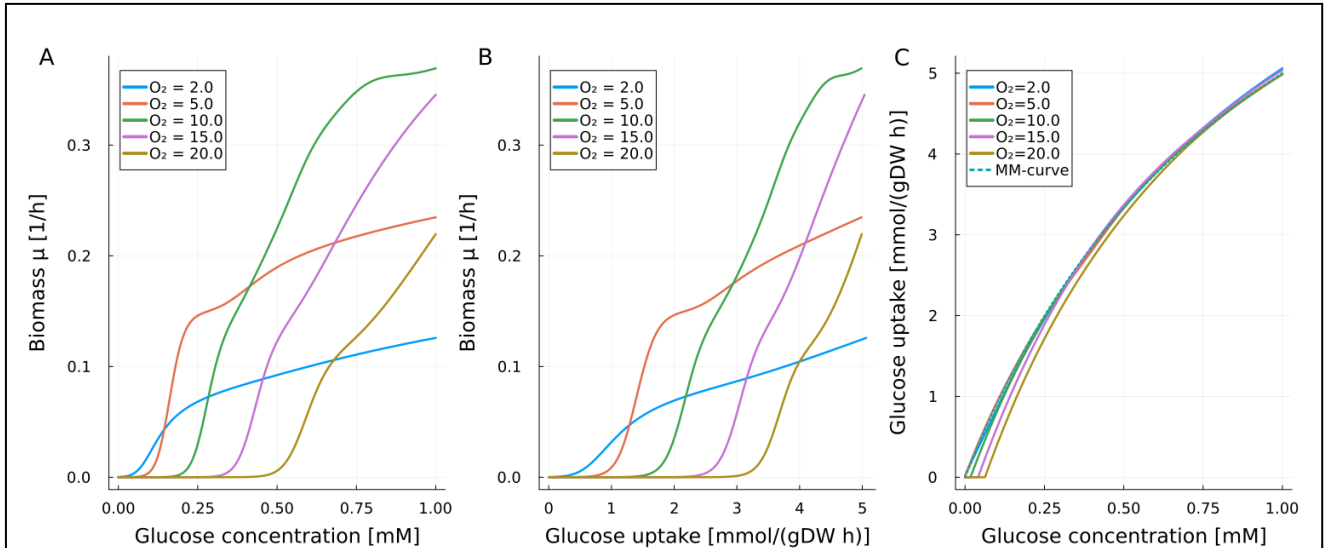

**Figure 2:**
Surrogate model predictions of intracellular metabolism for the *E. coli* genome scale metabolic model. Biomass growth rate as a function of extracellular glucose concentration (A) for different fixed oxygen uptake rates, with growth decreasing as glucose availability is reduced. Biomass growth rate as a function of glucose uptake rate (B) at fixed oxygen uptake rate. Glucose uptake rate as a function of extracellular glucose concentration (C), with uptake decreasing towards zero at low glucose concentrations. The dotted line in (C) shows a Michaelis-Menten curve for comparison with $V_{max} = 10$ and $K_m = 1.0$.

## Reactor modeling

The reactor model is modelled as a fed-batch system.

$$\dot{V} = e^T F_{in} \quad (2a)$$
$$\dot{n}_e = C_{in} F_{in} + R^*_{ext}(c_e)V, \quad c_e = \frac{n_e}{V} \quad (2b)$$
$$\dot{n}_X = R^*_X(c_e)V, \quad c_X = \frac{n_X}{V} \quad (2c)$$
$$R^*_{ext}(c_e) = v^T_{ext} r^*(c_e) \quad (2d)$$
$$R^*_X(c_e) = v^T_X r^*(c_e) \quad (2e)$$
$$r^*(c_e) = \mu^*(c_e) c_X \quad (2f)$$

$V$ is the volume, $F_{in}$ is the inlet flow rate, $n_e$ is the mole amount of extracellular species, $C_{in}$ denotes inlet concentration. The vector $R^*_{ext}(c_e)$ represents the extracellular production and consumption rates, obtained from the intracellular model. $n_X$ is the amount of biomass, and $R^*_X(c_e)$ is the production rate of biomass. $v_{ext}$ and $v_X$ are the stoichiometric matrices, $r^*(c_e)$ is the reaction rate vector, and $\mu^*(c_e)$ is the flux per unit of biomass and is a function of the extracellular concentration.

## Constraint-based modeling

Flux balance analysis assumes that intracellular metabolism operates at steady-state and computes a flux distribution consistent with stoichiometric constraints and the specified objective [3,12]. To account for changing extracellular conditions, uptake rates are fixed as functions of extracellular concentrations, denoted $\bar{\mu}_{in}(c_e)$. Embedding FBA within the reactor model yields a dFBA formulation [5,12]. Intracellular reactions are assumed to equilibrate instantaneously relative to reactor-scale dynamics [5,7]. The FBA problem is reformulated using positive forward and reverse fluxes to ensure directional consistency.

$$\mu^*(c_e) = \mu^* = \min_\mu \quad c^T(\mu^+ - \mu^-) \quad (3a)$$
$$s.t. \quad R_c = v_c^T(\mu^+ - \mu^-) = 0 \quad (3b)$$
$$0 \leq \mu^+ \leq \mu^+_{max} \quad (3c)$$
$$0 \leq \mu^- \leq \mu^-_{max} \quad (3d)$$
$$\mu^+_{in} - \mu^-_{in} = \bar{\mu}_{in}(c_e) \quad (3e)$$

$\mu$ is the vector of fluxes, decomposed into forward and reverse components, $\mu^+$ and $\mu^-$. The vector $c$ defines the linear objective function, the matrix $v_c$ is the intracellular stoichiometric matrix, and the constraint $R_c = 0$ enforces steady-state balances for intracellular metabolites. The vectors $\mu^+_{max}$ and $\mu^-_{max}$ define the upper bounds for the forward and reverse fluxes. The FBA solution is generally non-unique, as the steady-state stoichiometric constraints define a high-dimensional feasible set [7, 13]. Linear optimization selects a single point from the optimal face, so parsimonious FBA (pFBA) is applied to identify flux distributions with minimal total flux, while maintaining the optimal growth rate $z^*_{FBA}$ [14]. The pFBA optimization problem is defined as,

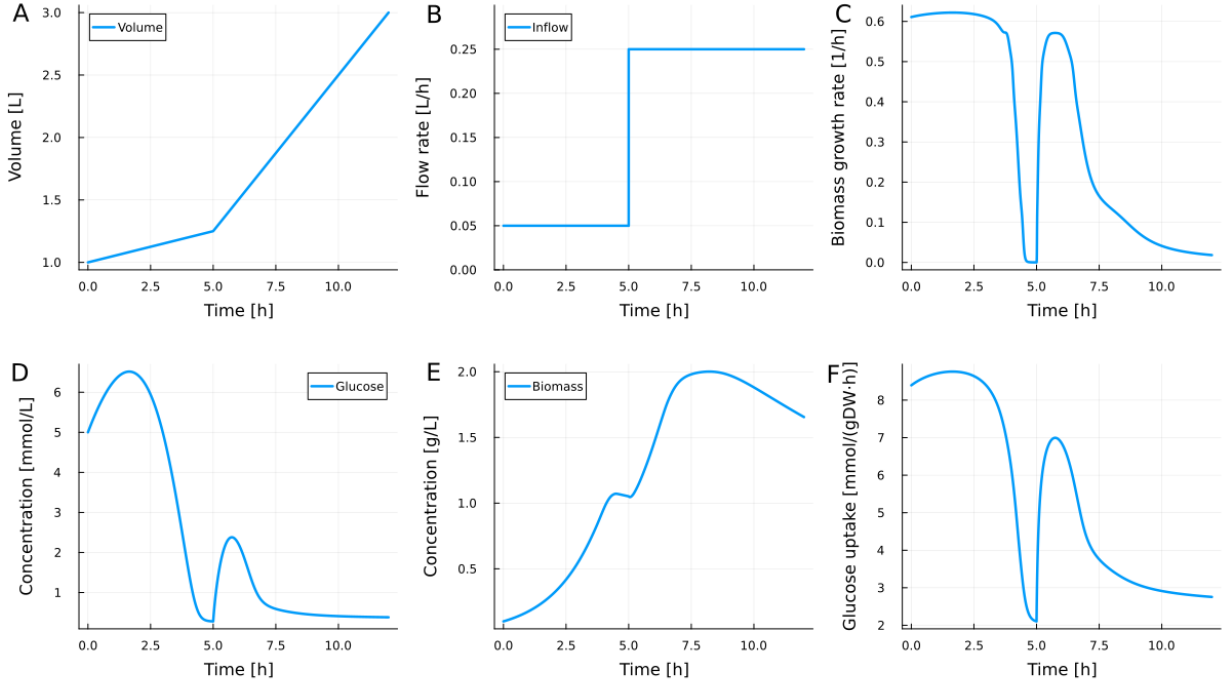

**Figure 3:**
Dynamic fed-batch simulation of *E. coli* using the proposed multi-scale modeling framework with a neural network surrogate model. The reactor volume (A) increases over time in response to the feed profile (B). Biomass growth rates (C) are initially high and subsequently decrease as extracellular glucose (D) is consumed. This results in biomass accumulation (E), while glucose uptake rates (F) decrease but remain non-zero due to the continuous glucose feed. Glucose uptake follows a Michaelis-Menten dependence on extracellular concentration. Biomass growth ceases before complete substrate depletion when substrate uptake is insufficient to sustain growth. Oxygen uptake is kept constant at 15 mmol/(gDW h).

$$\mu^*(c_e) = \mu^* = \min_{\mu} \sum_{i=1}^{n_\mu} \mu_i^+ + \mu_i^- \quad (4a)$$

$$s.t. \quad R_c = v_c^T(\mu^+ - \mu^-) = 0 \quad (4b)$$

$$0 \leq \mu^+ \leq \mu_{max}^+ \quad (4c)$$

$$0 \leq \mu^- \leq \mu_{max}^- \quad (4d)$$

$$\mu_{in}^+ - \mu_{in}^- = \bar{\mu}_{in}(c_e) \quad (4e)$$

$$\mu_X = z_{FBA}^* \quad (4f)$$

$n_\mu$ is the total number of reactions, and $\mu_X$ is the biomass growth flux.

### Surrogate modeling

During dynamic simulation, substrate depletion may render the linear program infeasible or lead to discontinuous flux trajectories [8]. To obtain a smooth representation of intracellular fluxes, a neural network surrogate is used to approximate the pFBA solution map [9,10]. The coupled system with the surrogate model can then be written as,

$$\dot{x} = f(x, y^*, u) \quad (5a)$$

$$y^*(x) = y^* = NN(\theta) \quad (5b)$$

$NN(\theta)$ is the trained neural network parameterized by $\theta$. The surrogate model is trained on sampled uptake rates and the corresponding optimal flux distributions obtained from pFBA. In regions where the original pFBA formulation becomes infeasible, feasibility is maintained by introducing slack variables [15]. The surrogate is trained to approximate the resulting relaxed solution map, with fluxes decreasing smoothly towards zero. The surrogate model thus provides a smooth approximation of the optimal flux distribution within the sampled training region.

Training data consist of non-negative parsimonious fluxes. The surrogate is implemented in Julia [16] using Flux.jl [17,18] and was trained using an NVIDIA H100 GPU and an AMD EPYC 9354 CPU. Training was performed on 40.401 pFBA solutions, each consisting of 349 intracellular fluxes. The pFBA solutions were generated from a uniform grid sampling over glucose and oxygen uptake rates from 0-20 mmol/(gDW h) with a step size of 0.1. The model was trained for approximately 15.000 epochs with early stopping.

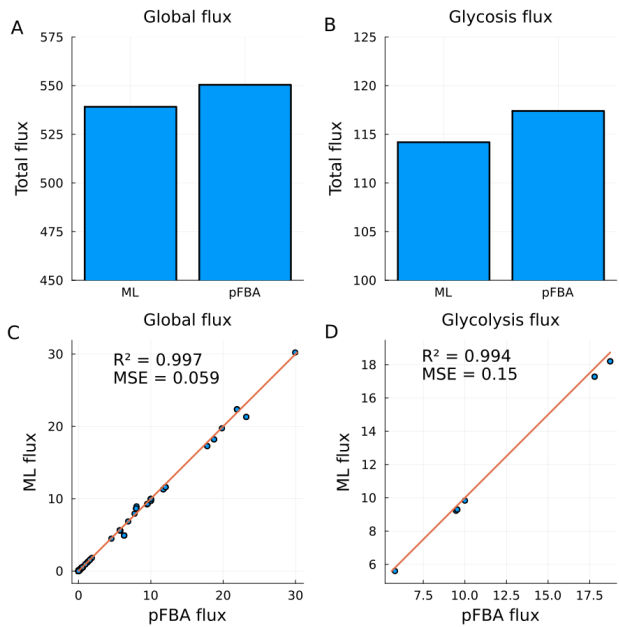

**Figure 4.** Comparison of surrogate and pFBA flux predictions using the *E. coli* genome-scale model. (A) Total absolute intracellular flux magnitude and (B) total glycolysis flux magnitude predicted by the surrogate and the pFBA at fixed glucose and oxygen uptake rates. (C-D) Per reaction flux comparison between surrogate and pFBA for the full network and the glycolysis pathway.

## RESULTS

Figure 2 shows surrogate model predictions of intracellular metabolism in *E. coli* under various conditions. Panel A presents how the biomass growth rate changes as a function of the extracellular glucose concentration. In the low-substrate regions shown, growth decreases as glucose availability decreases and approaches zero. Higher oxygen uptake rates (15 and 20 mmol/(gDW h)) have not plateaued and can sustain higher biomass growth at higher glucose concentrations. Panel B shows the predicted biomass growth rate as a function of glucose uptake rate at a fixed oxygen uptake rate. Panel C shows the predicted glucose uptake rate as a function of extracellular glucose concentration. Uptake decreases smoothly towards zero as glucose concentration decreases and approximates the Michaelis-Menten kinetic equation (dotted line for comparison). When there is not enough glucose available for a fixed oxygen uptake, the glucose uptake rate smoothly approaches zero. Under fermentation conditions, neither the substrate nor oxygen uptake is fixed, but instead adapts to an optimum at the conditions the cell is in.

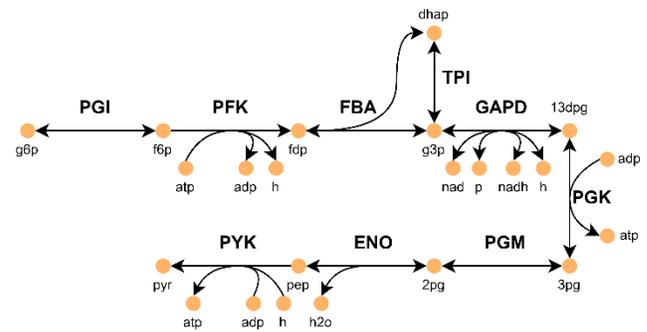

**Figure 5.** Graphical overview of the condensed glycolysis reactions used in Figure 4 (B, D). The enzyme abbreviations are in bold, and the yellow dots are intracellular species.

Figure 3 shows a dynamic fed-batch simulation of *E. coli* using the multi-scale modeling framework with the neural network surrogate. Panels A and B show the reactor volume and inflow, with a step increase in the inflow rate after 5 hours. Panel C shows the growth rate predicted by the surrogate model, while panels D and E show the glucose and biomass concentration in the bioreactor. Initially, the biomass concentration increases at a maximum rate under a fixed oxygen uptake. As the glucose concentration decreases, both the growth rate and glucose uptake shown in panel F decline. Following the step change in inflow, the reactor volume and glucose concentration increase, thereby increasing substrate availability. As a result, the glucose uptake, biomass growth rate, and biomass concentration increase. Once biomass accumulation is sufficient to take up more substrate than is supplied to the bioreactor, glucose concentration and biomass growth rate decrease. Here, the biomass concentration decreases due to dilution, while the total amount of biomass in the reactor continues to increase until the biomass growth rate is zero. At the step change, the non-surrogate model would have encountered an infeasible solution, and the simulation would have been terminated before additional feed [7].

Surrogate and pFBA flux predictions are compared in Figure 4 at glucose and oxygen uptake rates of 10 mmol/(gDW h). Panel A compares the total absolute intracellular flux, summed across all reactions, and the surrogate, which slightly underpredicts the sum of fluxes. Panel B shows the total flux related to glycolysis, specifically, the glycolysis reactions in Figure 5. In both cases, the surrogate reproduces the magnitude of the fluxes. Panels C and D show the parity plot comparing per-reaction flux predictions. Panel C includes all reactions, while Panel D consists of the glycolysis reactions in Figure 5. The low residuals and linear correlation between the surrogate and the pFBA flux predictions indicate a good fit for individual reactions.

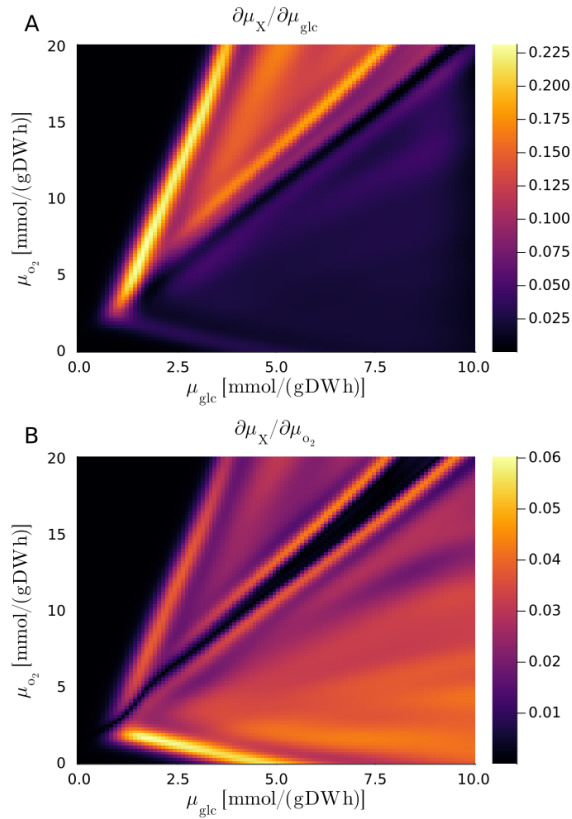

**Figure 6.** Sensitivity of the surrogate predicted biomass production rate $\mu_X$ with respect to glucose uptake, $\mu_{glc}$, and oxygen uptake, $\mu_{o_2}$.

It should be noted that the uptake rates used in Figures 3 and 4 are included in the surrogate training data. The surrogate model should be applied only within the region of the input space covered by the training data, as it does not explicitly enforce mass-balance constraints or upper and lower bounds. Instead, the surrogate model reproduces empirical relationships observed during training.

Figure 6 shows sensitivity measures for the surrogate intracellular metabolism across the glucose and oxygen uptake range. Panel A shows the partial derivative of the biomass production rate with respect to glucose uptake, $\partial \mu_X / \partial \mu_{glc}$. High sensitivity is observed along a narrow diagonal region, where concurrent increases in glucose and oxygen uptake lead to increased biomass production. Sensitivity decreases outside this region and approaches zero in low-uptake or oxygen-limited regions. Panel B shows the partial derivative of the biomass production rate with respect to oxygen uptake, $\partial \mu_X / \partial \mu_{o_2}$. Similar regions are observed, however, the overall magnitude of the oxygen sensitivity is lower and more broadly distributed than that of glucose sensitivity.

## DISCUSSION

In the iCH360 model, there is a minimum energetic requirement through the non-growth-associated maintenance reaction [3,13]. At sufficiently low substrate uptake rates, this requirement cannot be met, and the pFBA problem becomes infeasible [8]. To have training samples, slack variables are introduced, and the surrogate constitutes a relaxed pFBA solution in these regions. The surrogate was trained using glucose and oxygen uptake rates as inputs, assuming Michaelis-Menten uptake kinetics without inhibition [1]. Mass-balance constraints and flux bounds are not explicitly enforced and are reflected only through the training data. Additional inputs, such as mineral uptake or alternative substrates, would require retraining [10,19]. However, this retraining is performed offline and does not affect computational cost during simulation.

For the iCH360 model, solving the pFBA problem requires approximately 2.33 ms in Julia, where two linear programs are solved sequentially using Gurobi at fixed glucose and oxygen uptake rates. A classical FBA formulation requires approximately 120 $\mu$s. In contrast, the surrogate model predicts via a forward pass consisting of matrix-vector multiplications and nonlinear activation functions, in approximately 60 $\mu$s. This is 39 times faster than the pFBA-based formulation, and twice as fast as the classical FBA formulation.

The genome-scale model contains 349 intracellular fluxes, all of which were included in the training and validation datasets. Deviations between surrogate and pFBA predictions were generally small. The largest discrepancy between the pFBA solution and the surrogate prediction was observed for the hydrogen exchange reaction. Here, an absolute difference of 1.75 mmol/(gDW h), approximately 8%, was observed at a fixed glucose and oxygen uptake rate of 10 mmol/(gDW h). Increasing training coverage, adding more data points, or employing more advanced machine learning methods may reduce discrepancies [10,19].

When implemented in Julia [16] using Flux.jl [17,18], automatic differentiation is used to compute Jacobians and Hessians of intracellular fluxes with respect to states and parameters. State and parameter sensitivities are directly accessible through the formulation as well as for Julia's differential equation solvers. These sensitivities may be used in gradient-based state estimation and control [9].

## CONCLUSION

This work extends dynamic FBA by replacing the embedded linear optimization problem with a surrogate model of intracellular metabolism. By approximating the parsimonious FBA solution, the surrogate enables dynamic simulations in operating regions, where the original optimization problem may become infeasible. The surrogate produces continuous intracellular flux trajectories and eliminates repeated optimization solves during simulation. Extension to additional substrates and operating conditions can be achieved through offline retraining. The surrogate provides a computationally efficient basis for integration into control frameworks to optimize fermentation processes.

## ACKNOWLEDGEMENTS

Peter E. Carstensen is funded by The Novo Nordisk Foundation Center for Biosustainability, Technical University of Denmark, NNF20CC0035580.

## AUTHOR IDENTIFIERS

Author ORCIDs:
Carstensen, PE: 0009-0008-9019-7607
Groves, T: 0000-0002-7109-3270
Nielsen, LK: 0000-0001-8191-3511
Krühne, U: 0000-0001-7774-7442
Gernaey, K: 0000-0002-0364-1773
Jørgensen, JB: 0000-0001-9799-2808